\documentclass[aps,prb,twocolumn,showpacs,showkeys,preprintnumbers,groupedaddress,amsmath,amssymb,]{revtex4-1}

\usepackage{graphicx}
\usepackage{dcolumn}
\usepackage{bm}
\usepackage[dvips]{color}

\begin{document}



\title{Electron-electron scatttering in  Sn-doped indium oxide thick films}


\author{Yu-Jie Zhang}
\author{Zhi-Qing Li}
\email[Electronic address: ]{zhiqingli@tju.edu.cn}
\affiliation{Tianjin Key Laboratory of Low Dimensional Materials Physics and
Preparing Technology, Department of Physics, Tianjin University, Tianjin 300072, China}
\author{Juhn-Jong Lin}
\email[Electronic address: ]{jjlin@mail.nctu.edu.tw}
\affiliation{NCTU-RIKEN Joint Research Laboratory and Institute of Physics, National Chiao Tung University, Hsinchu 30010, Taiwan\\
and Department of Electrophysics, National Chiao Tung University, Hsinchu 30010, Taiwan}

\date{\today}

\begin{abstract}

We have measured the low-field  magnetoresistances (MRs) of a series of Sn-doped indium oxide thick films in the temperature $T$ range 4--35 K. The electron dephasing rate $1/\tau_{\varphi}$ as a function of $T$ for each film was extracted by comparing the MR data with the three-dimensional (3D) weak-localization theoretical predictions. We found that the extracted $1/\tau_{\varphi}$ varies linearly with $T^{3/2}$. Furthermore, at a given $T$, $1/\tau_{\varphi}$ varies linearly with $k_F^{-5/2}l^{-3/2}$, where $k_{F}$ is the Fermi wavenumber, and $l$ is the electron elastic mean free path. These features are well explained in terms of the small-energy-transfer electron-electron scattering time in 3D disordered conductors. This electron dephasing mechanism dominates over the electron-phonon ($e$-ph) scattering process because the carrier concentrations in our films are $\sim$ 3 orders of magnitude lower than those in typical metals, which resulted in a greatly suppressed $e$-ph relaxation rate.

\end{abstract}

\pacs{72.15.Rn, 73.20.Fz, 73.23.-b, 72.15.Qm}

\maketitle

\section{Introduction}

The problem of the electron dephasing  processes in disordered conductors has long been of fundamental importance and great interest. In general, the responsible electron dephasing mechanisms are determined by the system dimensionality, the level of disorder, and the measurement temperature $T$. \cite{Zyj1,zyj3,zyj4} In three-dimensional (3D) weakly disordered metals, electron-phonon (\emph{e}-ph) scattering is often the dominant dephasing mechanism. \cite{zyj4,zyj6,Zhong-prl98} While in lower dimensions, electron-electron (\emph{e}-\emph{e}) scattering is the major dephasing process, which gives rise to a phase relaxation rate of $1/\tau_{ee}\propto T$ in two dimensions (2D) \cite{Zyj1,zyj4,zyj8} and $1/\tau_{ee}\propto T^{2/3}$  in one dimension. \cite{Zyj1,zyj4,zyj9} In 1974, Schmid had already theoretically investigated the possible influence of disorder on the \emph{e}-\emph{e} scattering in 3D conductors. \cite{zyj10} He found that the diffusive electron dynamics should enhance the scattering strength, and thus the total \emph{e}-\emph{e} scattering rate could be written as \cite{zyj10}
\begin{equation}\label{Eq.(ee3d)}
\frac{1}{\tau_{ee}}=\frac{\pi}{8}\frac{(k_B T)^2}{\hbar E_F}+\frac{\sqrt{3}}{2\hbar\sqrt{E_F}}\left( \frac{k_B T}{k_F l}\right)^{3/2},
\end{equation}
where $k_B$ is Boltzmann constant, $E_F$ ($k_F$) is the Fermi energy (wavenumber), $\hbar$ is the Planck constant divided by $2\pi$, and $l$ is the electron elastic mean free path. A similar expression has also been derived by Altshuler and Aronov. \cite{zyj11} The first term on the right-hand side of Eq.~(\ref{Eq.(ee3d)}) is the \emph{e}-\emph{e} scattering rate in a perfect, periodic potential,  while the second term is the enhanced contribution due to the introduction of imperfections (defects, impurities, etc.). Microscopically, the second term stands for the small-energy-transfer \emph{e}-\emph{e} scattering process and is responsible at $k_BT < \hbar/\tau_e$, while the first term represents the large-energy-transfer process and would dominate at $k_BT > \hbar/\tau_e$, where $\tau_e$ is the electron elastic mean-free time. \cite{zyj11,zyj23} As mentioned, the scattering rate predicated in Eq.~(\ref{Eq.(ee3d)}) is much weaker than the \emph{e}-ph scattering rate in typical 3D disordered metals. Thus, the prediction of Eq.~(\ref{Eq.(ee3d)}) has not been fully experimentally tested thus far, although a qualitative (but not quantitative) $T^{3/2}$ temperature dependence of $1/\tau_{ee}$ has occasionally been reported in literature. \cite{zyj12,zyj14,zyj15,zyj16} To provide a convincing test of the validity of the second term of Eq.~(\ref{Eq.(ee3d)}), in addition to the temperature dependence $1/\tau_{ee} \propto T^{3/2}$, the variation on carrier concentration ($k_F$) and disorder ($l$), i.e., $1/\tau_{ee} \propto k_F^{-5/2}l^{-3/2}$, is of critical importance.

Tin-doped indium oxide (ITO) is the most widely used transparent conducting oxide in current optoelectronic devices. In terms of the advantages for fundamental studies, the ITO material possesses a unique {\em free-electron-like} energy bandstructure. \cite{Mryasov-prb01,Odaka-jjap01,Wu-jap10} The electron concentrations, $n$, are $\sim$ 2 to 3 orders of magnitude lower than those in typical metals, and essentially independent of temperature. \cite{Wu-jap10,zyj17,zyj18,zyj19,zyj20} Recently, Wu and coworkers have studied the electron dephasing processes in 2D ITO thin films.\cite{zyj22} They found that the $e$-$e$ scattering dominated the dephasing in a wide $T$ range from liquid-helium temperatures up to nearly 100 K. In 3D disordered conductors, we notice that the $e$-ph relaxation rate, $1/\tau_{e\text{-}\text{ph}}$, scales essentially linearly with $n$ (Refs. \onlinecite{zyj6}, \onlinecite{zyj22} and \onlinecite{zyj34}), while the second term in Eq.~(\ref{Eq.(ee3d)})  predicts $1/\tau_{ee} \propto E_F^{-1/2}k_F^{-3/2}$. Approximately, we may write $1/\tau_{ee}  \propto k_F^{-5/2} \propto 1/n$ and estimate the relative dephasing strength to vary roughly as $(1/\tau_{ee})/(1/\tau_{ep}) \propto 1/n^2$. In other words, one can expect the $e$-$e$ scattering to dominate the total dephasing in those 3D systems having sufficiently low values of $n$. Indeed, as to be shown in this paper, ITO thick films fulfill this criterion and can be used to manifest the small-energy-transfer $e$-$e$ scattering rate predicted in Eq.~(\ref{Eq.(ee3d)}). Moreover, since ITO possesses free-electron-like characteristics as aforementioned, the relevant parameters ($E_F$, $k_F$, $l$, etc.) can be faithfully evaluated through electrical-transport measurements. The theoretical evaluation of Eq.~(\ref{Eq.(ee3d)}) can thus be known to a high degree of accuracy. In this paper, we report our experimental confirmation for the numerical prediction of Eq.~(\ref{Eq.(ee3d)}), by studying a series of ITO thick films in a wide $T$ range 4--35 K. Our ITO thick films have values of $n \sim 2 \times 10^{20}$ cm$^{-3}$ (see Table~\ref{TableLi}), as compared with $n \sim 1 \times 10^{23}$ cm$^{-3}$ in typical metals. \cite{Kittel}

\begin{table*}
\caption{\label{TableLi} Parameters for four representative ITO thick films. O$_p$ and $T_{s}$ are the percentage of oxygen and substrate temperature, respectively, during deposition,  $\rho$ is resistivity, $\emph{n}$ is carrier concentration, and $\emph{D}$ is diffusion constant. $1/\tau^{0}_{\varphi}$ and $A_{ee}$ are defined in Eq.~(\ref{Eq-Fittao}). $A^{th}_{ee}$ is the theoretical \emph{e}-\emph{e} scattering strength predicted by the second term of Eq.~(\ref{Eq.(ee3d)}).}
\begin{ruledtabular}
\begin{center}
\begin{tabular}{cccccccccccccc}
 Film  & $O_{p}$   & $T_{s}$   & thickness & $\rho(10\,\text{K})$     & $n(10\,\text{K})$ & $\emph{D}(10\,\text{K})$  & $\emph{k}_{F}\emph{l}$       &$1$/$\tau^{\text{0}}_{\text{$\varphi$}}$ & $A_{ee}$ & $A^{th}_{\text{ee}}$\\
           &   (\%)     & (K)   & ($\mu$m) & (m$\Omega$ cm)    &($10^{20}$ cm$^{-3}$)& (cm$^{2}$$/$s)  & &  ($10^{9}$ s$^{-1}$)   &   (K$^{-3/2}$ s$^{-1}$)  & (K$^{-3/2}$ s$^{-1}$) \\  \hline
1  &    0.8    & 670 & 1.39 &1.93       &2.3             &1.3  &  3.3     &           1.6          &2.3$\times$$10^8$   &2.9$\times$$10^8$  \\
2  &    1.0    & 630 & 1.20 &2.98       &1.7             &0.92 &  2.4     &           3.3          &2.8$\times$$10^8$   &5.4$\times$$10^8$  \\
3  &    1.2    & 720 & 1.45 &1.88       &3.4             &1.2  &  3.0     &           2.8          &2.1$\times$$10^8$   &6.7$\times$$10^8$  \\
4  &    4.0    & 670 & 1.35 &2.49       &2.0             &1.0  &  2.7     &           2.4          &2.5$\times$$10^8$   &4.2$\times$$10^8$  \\
\end{tabular}
\end{center}
\end{ruledtabular}
\end{table*}

\section{Experimental method}

Our ITO thick films were deposited on glass substrates by the standard rf sputtering method. A commercial Sn-doped In$_2$O$_3$ target (99.99\% purity, the atomic ratio of Sn to In being 1:9) was used as the sputtering source. The base pressure of the vacuum chamber was $8$$\times$$10^{-5}$ Pa, and the sputtering deposition was carried out in a mixture of argon and oxygen (99.999\% purity) atmosphere with a pressure of 0.6 Pa. During the deposition process, the percentage of oxygen O$_p$, together with the substrate temperature $T_s$, was varied to ``tune" the carrier (electron) concentration and the amount of disorder. Two series of ITO thick films were fabricated. In the first series, $T_s$ was fixed at 630, 650, 670, 690, or 720 K; and for each $T_s$, O$_p$ was kept at 0.8\%, 1.0\%, or 1.2\%. In the second series, $T_s$ was fixed at 670 K, and O$_p$ was kept at 1.4\%, 1.8\%, 2.0\%, or 4.0\%. Altogether, 19 samples with different amounts of disorder and carrier concentrations were deposited. Hall-bar-shaped samples (1-mm wide and 1-cm long) were defined by using mechanical masks \cite{zyj22} and used for electrical-transport measurements.

The thicknesses of the films were determined by a surface profiler (Dektak, 6M). The films were all at least 1 $\mu$m thick to ensure that they were 3D with respect to the weak-localization (WL) and \emph{e}-\emph{e} interaction effects. The structures of the films were determined by a Rigaku x-ray diffractometer (D/max-2500v/pc) with Cu $K_\alpha$ radiation. The measurements indicated that the films were single phased with a cubic bixbyite structure characteristic to that of the undoped In$_2$O$_3$. The observed strongest and second strongest diffraction peaks corresponded to the (400) and the (800) planes, respectively. The intensities of other diffraction peaks were less than one thirtieth that of the (400) peak. Thus, the preferred growth orientation was along the $[100]$ direction. The low-magnetic-field magnetoresistance (MRs) were measured on a physical property measurement system (PPMS-6000, Quantum Design). The magnetic fields were always applied perpendicular to the film plane. Hall effect measurements were also performed to determine the values of $n$.

\section{Results and discussion}

\begin{figure}[htp]
\begin{center}
\includegraphics[scale=1.06]{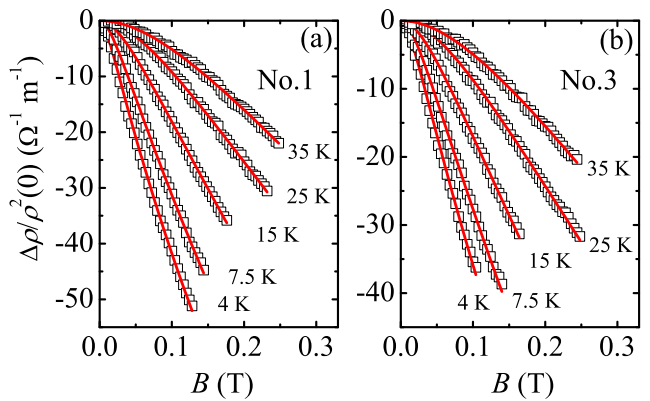}
\caption{(Color online) Normalized magnetoresistivity $\Delta \rho(B)/\rho^{2}(0)$ as a function of magnetic field at several temperatures for two ITO thick films, as indicated. The magnetic field was applied perpendicular to the film plane. The solid curves are least-squares fits to Eq.~(\ref{Eq-WL}).}
\label{MR}
\end{center}
\end{figure}

Figure~{\ref{MR}} shows the measured normalized magnetoresistivities, $\Delta \rho(B)/\rho^{2}(0)$ $=$ $[\rho(B)$ $-$ $\rho(0)]$$/\rho^{2}(0)$, as a function of magnetic field $B$ at several temperatures for two representative ITO films, as indicated. Note that all the other films revealed similar behavior. The MRs are negative at all temperatures and their magnitudes decrease with increasing $T$. In our films, we evaluate the product $k_{F}l \approx$ 1.7--3.5, and thus the WL effect must be relatively pronounced. The MR in the 3D WL effect is predicted to be \cite{zyj25,Wu-prb94} (with the function $f_3$ defined in Ref. \onlinecite{f3}),
\begin{equation}\label{Eq-WL}
\frac{\Delta \rho(B)}{\rho^{2}(0)}=\frac{e^2}{2 \pi^2 \hbar}\sqrt{\frac{eB}{\hbar}}\left[\frac{1}{2} f_{3}\left(\frac{B}{B_ \varphi}\right)-\frac{3}{2}f_{3}\left(\frac{B}{B_2}\right)\right] \,,
\end{equation}
where $B_\varphi = B_\text{i} + B_0$ and $B_2 = B_\text{i} + {B_0}/3 + 4B_{\text{so}}/3 = B_\varphi + 4 B^\ast_{\text{so}}/3$. The characteristic field $B_j$ is related to the characteristic scattering time $\tau_j$ through the relation $B_j = \hbar/4eD\tau_j$, where $j=\text{i}$, so and 0 refer to the inelastic, spin-orbit, and $T$-independent scattering times, $e$ is the elementary charge, and $D$ is the diffusion constant.

Our MR data are least-squares fitted to Eq.~(\ref{Eq-WL}). We found that, in all films, $B^\ast_\text{so}$ is $\sim$ 2 orders of magnitude smaller than $B_{\varphi}$(4 K). That is, effectively, $B^\ast_\text{so}$ may be set to zero and $B_\varphi$ becomes the sole adjustable parameter. Indeed, the negative MRs at all measurement temperatures already suggest that the spin-orbit scattering must be negligibly weak. The solid curves in Fig.~\ref{MR} represent the theoretical predications of Eq.~(\ref{Eq-WL}). Clearly, the MR data can be well described by the WL theory. We obtain the electron dephasing length $L_{\varphi} = \sqrt{D\tau_{\varphi}}$ at 4 K varying from $\sim$ 110 to $\sim$ 190 nm in our films. Thus, our thick films are 3D with regard to the WL effect.

\begin{figure}[htp]
\begin{center}
\includegraphics[scale=1.2]{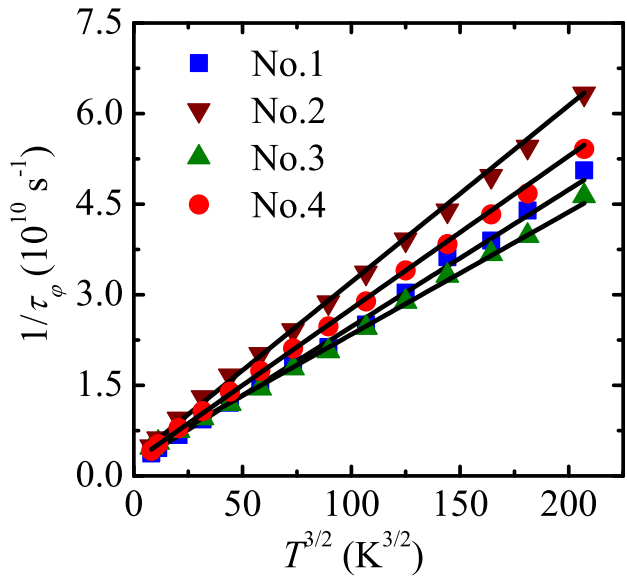}
\caption{(Color online) Electron dephasing rate $1/\tau_{\varphi}$ as a function of temperature for four ITO thick films, as indicated. The solid curves
are least-squares fits to Eq.~(\ref{Eq-Fittao}).}
\label{tao-T}
\end{center}
\end{figure}

Figure~\ref{tao-T} plots our extracted electron dephasing rate $1/\tau_{\varphi}$ as a function of $T^{3/2}$ for four representative films, as indicated. Note that the product $k_Fl \approx$ 1.7--3.5 in our films, and hence the first term in Eq.~(\ref{Eq.(ee3d)}) is irrelevant in the following discussion. \cite{clean-term} Inspection of Fig.~\ref{tao-T} indicates that $1/\tau_{\varphi}$ varies linearly with $T^{3/2}$ in the wide measurement temperature range 4--35 K. We compare our $1/\tau_\varphi$ data to the following expression
\begin{equation}\label{Eq-Fittao}
\frac{1}{\tau_{\varphi}}=\frac{1}{\tau^{0}_{\varphi}}+A_{ee}T^{3/2} \,,
\end{equation}
where the first term on the right-hand side stands for a $T$-independent (or weakly $T$ dependent) contribution, \cite{zyj29,zyj30,zyj31} and the second term stands for the 3D small-energy-transfer \emph{e}-\emph{e} scattering rate. Our least-squares fitted values of $1/\tau^{0}_{\varphi}$ and $A_{ee}$ are listed in Table~{\ref{TableLi}}. Also listed in Table~{\ref{TableLi}} are the corresponding theoretical values $A_{ee}^{th} = (\sqrt{3}/2 \hbar \sqrt{E_F})(k_B/k_Fl)^{3/2}$.  Inspection of Table~{\ref{TableLi}} indicates that our experimental values of $A_{ee}$ are within a factor of $\sim$ 3 of $A_{ee}^{th}$. This degree of agreement is satisfactory. Similar degree of agreement was found for all the other films studied in this work. We mention that, in this study, we have intentionally focused on $T \geq$ 4 K to minimize the influence of $1/\tau_\varphi^0$ on extracting $A_{ee}$.

\begin{figure}[htp]
\begin{center}
\includegraphics[scale=1.2]{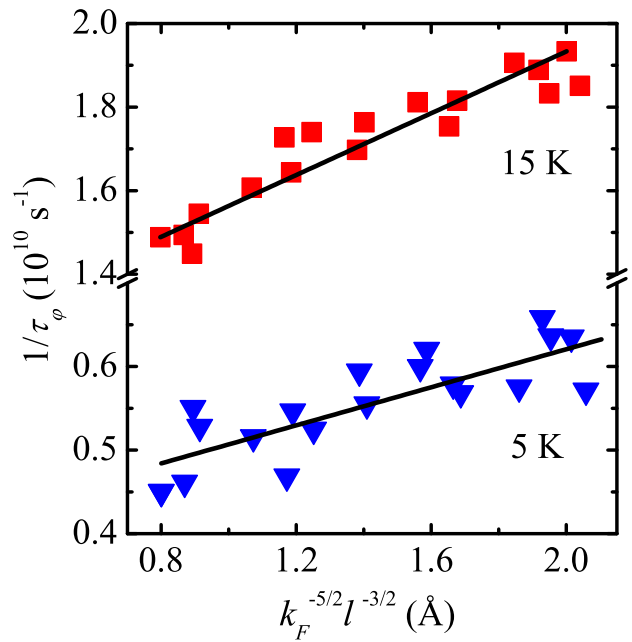}
\caption{(Color online) Electron dephasing rate $1/\tau_{\varphi}$ as a function of $k_F^{-5/2}l^{-3/2}$ at 5 and 15 K, as indicated. The solid lines are linear fits to the experimental data.}
\label{kfl}
\end{center}
\end{figure}

Figure~{\ref{kfl}} shows a plot of the variation of our extracted $1/\tau_{\varphi}$ with $k_F^{-5/2}l^{-3/2}$ at two representative $T$ values, as indicated. This figure clearly reveals a linear variation, as it should be according to the second term of Eq.~(\ref{Eq.(ee3d)}). Our least-squares fits give slopes of $\simeq 1.2 \times 10^{19}$ m$^{-1}$ at 5 K and $\simeq 3.7 \times 10^{19}$ m$^{-1}$ at 15 K. The theoretical slope can be expressed as $(1.22 \sqrt{m^\ast} / \hbar^2) (k_B T)^{3/2}$ in the free-electron model, where the effective mass $m^\ast = 0.4\, m$ ($m$ being the free electron mass). \cite{zyj17} We calculate the theoretical slopes to be $\simeq 3.8 \times 10^{19}$ and $\simeq 2.0 \times 10^{20}$ m$^{-1}$ at 5 and 15 K, respectively. These values agree with the experimental values to within a factor of $\sim$ 5. Thus, in addition to the temperature dependence, our $1/\tau_\varphi$ data concerning the carrier concentration ($k_F$) and disorder ($l$) dependence fully support the Schmid-Altshuler-Aronov theory of 3D small-energy-transfer $e$-$e$ scattering time in disordered conductors. \cite{zyj10,zyj11}

Finally, we estimate the $e$-ph relaxation rate in our ITO thick films. It has recently been established that the electron scattering by transverse vibrations of defects dominate the \emph{e}-ph relaxation in the quasiballistic limit of $q_{T}l > 1$, where $q_T$ is the wavenumber of a thermal phonon. \cite{zyj6,zyj34} The relaxation rate is predicted to be \cite{zyj32,zyj34}
\begin{equation}\label{Eq-ep}
\frac{1}{\tau_{e\text{-}t,\text{ph}}}=\frac{3\pi^2 k_B^2 \beta_t}{(p_F u_t)(p_F l)}T^2 = A_{e\text{-}t,\text{ph}}T^2 \,,
\end{equation}
where $\beta_t = (2E_F/3)^2N(E_F)/(2\rho_m u_t^2)$ is the electron--transverse phonon coupling constant,  $p_F$ is the Fermi momentum, $u_t$  is the transverse sound velocity, $\rho_m$ is the mass density, and $N(E_F)$ is the electronic density of states  at the Fermi level. In the ITO material, $u_t \approx 2400$ m$/$s,  \cite{Wittkowski01} and the typical values of $l$ in our films are $\approx$ 1.5 nm. We evaluate the product $q_T l \approx k_B Tl/\hbar u_t \approx 0.09\, T$, where $T$ is in K. Thus, above about 10 K, our films lie in the quasiballistic limit. Substituting $\rho_m \approx$ 7100 kg$/$m$^{3}$ (Ref. \onlinecite{Wittkowski01}) and the relevant electronic parameters into Eq.~(\ref{Eq-ep}), we obtain the coupling strength $A_{e\text{-}t,\text{ph}} \sim 4 \times 10^6$ K$^{-2}$ s$^{-1}$. Thus, $1/\tau_{e\text{-ph}}$ is still about one order of magnitude smaller than $1/\tau_{ee}$ even at $T$ = 35 K. The smallness of $1/\tau_{e\text{-ph}}$ in ITO makes feasible our experimental observation of the 3D small-energy-transfer \emph{e}-\emph{e} scattering time in the wide $T$ range 4--35 K.

\section{Conclusion}

We have investigated the electron dephasing mechanisms in ITO thick films in a wide temperature range 4-35 K. Our electron dephasing times were extracted from the weak-localization magnetoresistance measurements. We obtained $1/\tau_\varphi \propto T^{3/2}$ and $1/\tau_\varphi \propto k_F^{-5/2}l^{-3/2}$, which can be well ascribed to the small-energy-transfer electron-electron scattering process in three-dimensional disordered systems. This observation was achieved because our ITO films possessed relatively low carrier concentrations which resulted in a greatly suppressed electron-phonon relaxation rate.

\begin{acknowledgments}

This work was supported by the NSF of China through Grant No. 11174216, Research Fund for the Doctoral Program of Higher Education through Grant
No. 20120032110065 (Z.Q.L.), and by the Taiwan National Science Council through Grant No. NSC 101-2120-M-009-005, and the MOE ATU Program (J.J.L.).

\end{acknowledgments}


\begin{thebibliography}{00}\label{sec:TeXbooks}

\bibitem{Zyj1} B. L. Altshuler, A. G. Aronov, and D. E. Khmelnitsky, J. Phys. C \textbf{15}, 7367 (1982).
\bibitem{zyj3} P. A. Lee and V. Ramakrishnan, Rev. Mod. Phys. \textbf{57}, 287 (1985).
\bibitem{zyj4} J. J. Lin and J. P. Bird, J. Phys. Condens. Matter \textbf{14}, R501 (2002).

\bibitem{zyj6} J. Rammer and A. Schmid, Phys. Rev. B \textbf{34}, 1352 (1986).

\bibitem{Zhong-prl98} Y. L. Zhong and J. J. Lin, Phys. Rev. Lett. \textbf{80}, 588 (1998).

\bibitem{zyj8} H. Fukuyama and E. Abrahams, Phys. Rev. B \textbf{27}, 5976 (1983).
\bibitem{zyj9} F. Pierre, A. B. Gougam, A. Anthore, H. Pothier, D. Esteve, and N. O. Birge, Phys. Rev. B \textbf{68}, 085413 (2003).
\bibitem{zyj10} A. Schmid, Z. Phys. \textbf{271}, 251 (1974).
\bibitem{zyj11} B. L. Altshuler and A. G. Aronov, JETP Lett. \textbf{30}, 482 (1979).
\bibitem{zyj23} B. L. Altshuler and A. G. Aronov, in \textit{Electron-Electron Interactions in Disordered Systems}, edited by A. L. Efros and M. Pollak (Elsevier, Amsterdam, 1985).
\bibitem{zyj12} Z. Ovadyahu, Phys. Rev. Lett. \textbf{52}, 569 (1984).
\bibitem{zyj14} A. Stolovits, A. Sherman, K. Ahn, and R. K. Kremer, Phys. Rev. B \textbf{62}, 10565 (2000).
\bibitem{zyj15} T. Andrearczyk, J. Jaroszy\'nski, G. Grabecki, T. Dietl, T. Fukumura, and M. Kawasaki, Phys. Rev. B \textbf{72}, 121309(R) (2005).
\bibitem{zyj16} T. Dietl, T. Andrearczyk, A. Lipi\'nska, M. Kiecana, M. Tay, and Y. Wu, Phys. Rev. B \textbf{76}, 155312 (2007).

\bibitem{Mryasov-prb01} O. N. Mryasov and A. J. Freeman, Phys. Rev. B \textbf{64}, 233111 (2001).

\bibitem{Odaka-jjap01} H. Odaka, Y. Shigesato, T. Murakami, and S. Iwata, Jpn. J. Appl. Phys. Part 1 \textbf{40}, 3231 (2001).

\bibitem{Wu-jap10} C. Y. Wu, T. V. Thanh, Y. F. Chen, J. K. Lee, and J. J. Lin, J. Appl. Phys. \textbf{108}, 123708 (2010).

\bibitem{zyj19} Z. Q. Li and J. J. Lin, J. Appl. Phys. \textbf{96}, 5918 (2004).

\bibitem{zyj17} I. Hamberg, C. G. Granqvist, K. F. Berggren, B. E. Sernelius, and L. Engstr\"{o}m, Phys. Rev. B \textbf{30}, 3240 (1984).
\bibitem{zyj18} N. Kikuchi, E. Kusano, H. Nanto, A. Kinbara, and H. Hosono, Vacuum \textbf{59}, 492 (2000).

\bibitem{zyj20} Y. J. Zhang, Z. Q. Li, and J. J. Lin, Phys. Rev. B \textbf{84}, 052202 (2011).

\bibitem{zyj22} C. Y. Wu, B. T. Lin, Y. J. Zhang, Z. Q. Li, and J. J. Lin, Phys. Rev. B \textbf{85}, 104204 (2012).

\bibitem{zyj34} A. Sergeev and V. Mitin, Phys. Rev. B \textbf{61}, 6041 (2000).

\bibitem{Kittel} C. Kittel, \textit{Introduction to Solid State Physics}, 8th ed. (Wiley, New York, 2005).

\bibitem{zyj25} H. Fukuyama and K. Hoshino, J. Phys. Soc. Jpn. \textbf{50}, 2131 (1981).

\bibitem{Wu-prb94} C. Y. Wu and J. J. Lin, Phys. Rev. B \textbf{50}, 385 (1994).

\bibitem{f3} The function $f_3$ is an infinite series and can be approximately expressed as $f_3(z) \approx 2 ( \sqrt{2+1/z} \, - \sqrt{1/z} ) - [ (2 + 1/z)^{-1/2} + (3/2 + 1/z)^{-1/2} ] + (1/48) (2.03 + 1/z)^{-3/2}$. See D. V. Baxter, R. Richter, M. L. Trudeau, R. W. Cochrane, and J. O. Strom-Olsen, J. Phys.  Paris \textbf{50}, 1673 (1989).

\bibitem{clean-term} The typical Fermi energy in our ITO thick films is $E_F \approx$ 0.6 eV. This value of $E_F$ leads to a clean-limit $e$-$e$ scattering rate of $\approx 5 \times 10^6 \,T^2$ s$^{-1}$ ($T$ in K) according to Eq.~(\ref{Eq.(ee3d)}). This scattering rate is about one order of magnitude smaller than the second term even at $T$ = 35 K.

\bibitem{zyj29} J. J. Lin and N. Giordano, Phys. Rev. B \textbf{35}, 1071 (1987).
\bibitem{zyj30} P. Mohanty, E. M. Q. Jariwala, and R. A. Webb, Phys. Rev. Lett. \textbf{78}, 3366 (1997).
\bibitem{zyj31} S. M. Huang, T. C. Lee, H. Akimoto, K. Kono, and J. J. Lin, Phys. Rev. Lett. \textbf{99}, 046601 (2007).
\bibitem{zyj32} Y. L. Zhong, A. Sergeev, C. D. Chen, and J. J. Lin, Phys. Rev. Lett. \textbf{104}, 206803 (2010).

\bibitem{Wittkowski01} T. Wittkowski, J. Jorzick, H. Seitz, B. Schroder, K. Jung, and B. Hillebrands, Thin Solid Films \textbf{398--399}, 465 (2001).

\end{thebibliography}

\end{document}